# The numerical sausage

L. Belardinelli and E. Onofri

*Dipartimento di Fisica dell'Università e Gruppo Collegato I.N.F.N.*
Parma, 43100 Italy

## Abstract

The renormalization group equation describing the evolution of the metric of the non linear sigma models poses some nice mathematical problems involving functional analysis, differential geometry and numerical analysis. We describe the techniques which allow a numerical study of the solutions in the case of a two-dimensional target space (deformation of the $O(3)$ $\sigma$–model). Our analysis shows that the so–called sausages define an attracting manifold in the $U(1)$–symmetric case, at one–loop level. The paper describes i) the known analytical solutions, ii) the spectral method which realizes the numerical integrator and allows to estimate the spectrum of zero–modes, iii) the solution of variational equations around the solutions, and finally iv) the algorithms which reconstruct the surface as embedded in $R^3$.

# 1 The renormalization group equation

The perturbative renormalization of the non linear $\sigma$–model [1] gives rise to a deformation of the metric according to the (one–loop) equation

$$\frac{\mathrm{d} g_{ij}}{\mathrm{d} t} = -\frac{1}{4\pi} R_{ij} + O(R^2) \tag{1}$$

This second order nonlinear partial differential equation has been studied in the simplest case of 2–dimensional target manifolds (Riemann surfaces) in Ref.[2]. A whole family of solutions is known for the topology of the sphere $S^2$ or for the torus. In this paper we consider the case of genus two ($S^2$) only, but our method are easily adapted to the toroidal case. As it is well known we can introduce local coordinates $\{y, \varphi\}$ in such a way that the metric is given by a conformal deformation of the flat metric:

$$g_{ij} = \exp(\phi)\, \delta_{ij} \tag{2}$$

and the RG equation reduces to a single (non–linear) partial differential equation

$$\frac{\partial \phi}{\partial t} = \beta(R) \tag{3}$$

the scalar curvature $R$ being

$$R = -\exp(-\phi)\, \boldsymbol{\partial}^{\mathbf{2}} \phi \tag{4}$$

with $\boldsymbol{\partial}^{\mathbf{2}} \phi = (\frac{\partial^2 \phi}{\partial y^2} + \frac{\partial^2 \phi}{\partial \varphi^2})$. A family of solutions in the special case of one-loop $\beta(x) = -x/4\pi$ has been presented in [2]. These can be obtained by introducing the *ansatz*

$$\phi(y, \varphi; t) = -\log\left(a(t) + b(t) f(y) + c(t) h(\varphi)\right) \tag{5}$$

The only solutions of this kind which can be extended to $t \to -\infty$ (ultraviolet limit) without encountering singularities and possess a residual U(1) symmetry are the so-called *sausage* solutions, parameterized by a single real constant $\nu$:

$$\phi(y) = -\log(a(t) + b(t)\cosh 2y)$$
$$a(t) = \tfrac{1}{2}\nu \coth(\nu(t_0-t)/(2\pi)), \quad b(t) = \tfrac{1}{2}\nu/\sinh(\nu(t_0-t)/(2\pi)) \tag{6}$$

These solutions have been shown to correctly describe the one-loop renormalized $\sigma$-model with U(1) symmetry [2]. To discuss the general solution of the RG equation we shall now construct a numerical integration algorithm. The first problem which one has to solve is that the conformal factor $\phi$ diverges as $-2|y|$ as $|y| \to \infty$, a general property which stems from the topology of the sphere (*i.e.* $\int R \, dy \, d\varphi/4\pi = 2 = 1/2(\frac{\partial \phi}{\partial y}(+\infty) - \frac{\partial \phi}{\partial y}(-\infty)))$. This fact has the undesirable effect of making the factor $\exp\{-\phi\}$ diverge, hence amplifying the numerical error of the differential term at large $|y|$. To overcome this difficulty we introduce a background field $\phi_0(y,t)$, and we consider the equation for the shifted field $\phi = \phi_0 + \eta$. The background field is conveniently chosen as the constant curvature solution

$$\phi_0(y,t) = \log\left(\frac{A(t)}{4\pi \cosh^2 y}\right) \tag{7}$$

which corresponds to $R(\phi_0) = 8\pi/A(t)$. According to the RG equation we get

$$\dot{A}(t)/A(t) = \beta(8\pi/A(t)) \tag{8}$$

It is easier to adopt A itself as the evolution parameter, or rather the new "time" $\tau$ defined by

$$A(t) = 4\pi \exp(-\tau) \tag{9}$$

while the original scale can be simply recovered by quadrature

$$t_0 - t = \int_{\tau_0}^{\tau} \frac{d\tau'}{\beta(2\exp(\tau'))} \tag{10}$$

When expressed in terms of the shifted field $\eta$ the RG equation reads [1]

$$\partial_\tau \eta = 1 - \beta(\tilde{R}(\eta))/\beta(2e^\tau)$$
$$\tilde{R}(\eta) = e^{\tau-\eta}(2 - \Delta_0 \eta) \tag{11}$$

Here $\Delta_0 \equiv \cosh^2 y \, \partial_y^2$ is the standard O(3) invariant Laplacean on the sphere. At this point it is convenient to adopt the standard spherical coordinates by letting $y = \log(\cot(\vartheta/2))$. To get a good accuracy in the evaluation of the Laplacean we apply the spectral method, that is we expand $\eta$ in Legendre polynomials—the eigenfunctions of $\Delta_0$. To do that we need some finite dimensional implementation of the standard expansion in Legendre polynomials. Since we have not been able to find such an algorithm in the literature, we had to develop our own.

---

[1] As a shorthand we set $\tilde{R}(\eta) \equiv R(\phi_0 + \eta)$.

## 1.1 The finite Legendre expansion

Let us pick some large integer $L$; let $x_j^{(L)}$ be the zeroes of the $L$-th Legendre polynomial $P_L(x)$. We adopt $\{x_j^{(L)}|j=1,\ldots,L\}$ as our finite grid.[2] the field $\eta(y,t)$ being sampled at the image points $y = \frac{1}{2}\log((1+x)/(1-x))$.

The finite Legendre expansion is then realized as follows:

$$\eta(x_j^{(L)}) = \sum_{l<L} \eta_l P_l(x_j^{(L)})$$

$$\eta_l = \int_{-1}^{1} dx \eta(x) P_l(x) \frac{2l+1}{2} \approx \sum_{j=1}^{L} \eta(x_j^{(L)}) \, P_l(x_j^{(L)}) \, \frac{2l+1}{2} \, w_j^{(L)} \qquad (12)$$

where $w_j^{(L)}$ are the Gaussian weights [3] for Legendre Polynomials. The finite Legendre expansion allows us to represent the Laplacean *exactly* on polynomials of degree $< L$ and *the inverse transform is also exact up to this degree*. The lack of a *fast* implementation (the analogous of FFT) limits our algorithm in practice to $L \sim 200$ on current workstations, but this proves to be adequate for our purposes.

While extensive tables for $\{x_j^{(L)}, w_j^{(L)}\}$ are available [4], it is far more convenient to generate them on line, applying a well–known algorithm based on the recursion relations of orthogonal polynomials; the idea is to build a (symmetric) companion matrix for $P_L$ in such a way that both the zeroes of $P_L$ and the Gaussian integration coefficients are evaluated simultaneously in the process of diagonalization (see Appendix A).

Having determined the finite transform $\mathcal{L}$ on the basis of Legendre polynomials, the action of the Laplacean $\mathbf{\Delta}_0$ is represented by a matrix $\mathcal{L}^{-1}\Lambda\mathcal{L}$, where $\Lambda$ is diagonal, with eigenvalues $\{-l(l+1)|l=0,\ldots,L-1\}$.

In terms of this representation it is quite easy to compute the spectrum of zero–modes of the field $\phi$, a problem considered in Ref.[2]. We have just to diagonalize the finite matrix $-\frac{1}{2}\mathbf{\Delta} + \frac{1}{8}\tilde{R}$, where $\mathbf{\Delta} = \exp\{\tau - \eta\}\mathbf{\Delta}_0$. The results agree with the previously computed ones for the sausages [5] (notice however that the present method is much simpler and the spectrum can be computed in parallel with the RG evolution).

## 1.2 The numerical integration of the RG equation

With the finite–Legendre–transform algorithm at hand, we can now consider the integration of Eq.11. We have implemented the algorithm in MATLAB which provides efficient routines of diagonalization and of adaptive-step integration (see App.B). The accuracy of the code has been tested on the known "sausage" solutions, attaining a typical maximal deviation of 1 part in $10^9$ over a time interval $-3 < \tau < 2$ and $\nu < .25$. The accuracy is limited essentially from the large eigenvalues of the Laplacean which grow as the square of the dimension of the finite grid. Moreover as $\nu$ and/or $-\tau$ increase the curvature tends to be confined at the extremities of the sausage, which requires finer and finer grids. Presently, with up to 200 points, we cannot go beyond $\tau \approx -4$ for $\nu \approx .25$, but there is no limit of principle.

---

[2]Solutions with reflection symmetry ($x \to -x$) can be studied by a restriction to even–order polynomials; our algorithm can be easily adapted to this case, with a factor of four gain in memory requirements.

The algorithm can now be applied to investigate the existence of *attracting manifolds* in the space of all metrics. A conjecture of Fateev and Zamolodchikov states that the sausages constitute a stable manifold and all other geometries converge to some sausage, parametrised by the real number $\nu$, in the infrared direction $\tau \to \infty$. This fact manifests itself quite clearly in our numerical data. We define a distance function by

$$\mathrm{dist}(\eta_1, \eta_2) = \int |\eta_1 - \eta_2|^2 \mathrm{d}x$$

and we measure the distance to the sausage hypersurface by

$$D(t) = \inf_\nu \mathrm{dist}(\eta(\tau), \eta_\nu(\tau))$$

The data show a clear exponential decay $D(t) \approx A \exp(-m\tau)$ with $m \approx 8$. For any given starting geometry we record the value of $\nu(t)$ where the infimum is reached; its limit as $\tau \to \infty$ gives a definition of *sausageness* of any given surface. For instance, given an ellipsoid with cylindrical symmetry and eccentricity $\epsilon$ we can measure $\nu(\epsilon)$, at one–loop order (see Tab.1, 2). At two loops sausages are not solutions any more; the characterization of the attracting manifold in this case is an open question, which can be explored by our algorithm.

## 2   Variational equations and stability

Another way to discuss the attracting nature of the sausages' manifold is given by the solution of Jacobi variational equations around the sausages' solutions. The linearized equations take on the following form:

$$\partial_\tau \chi = -\frac{\beta'(\tilde{R}(\eta))}{\beta(2e^\tau)} \frac{\delta \tilde{R}}{\delta \eta} \chi \equiv H(\eta, \tau) \chi \tag{13}$$

where

$$H(\eta, \tau) = \frac{\beta'(\tilde{R}(\eta))}{\beta(2e^\tau)} \left( e^{\tau - \eta} \mathbf{\Delta}_0 + \tilde{R}(\eta) \right) \tag{14}$$

The spectrum of $H$ is not a priori of much significance, since the evolution equations are time–dependent. However, if we rely on the adiabatic approximation, the spectrum is directly related to stability. Applying the discrete–Legendre–transform the spectrum of $H$ is easily reduced to that of a Hermitian matrix; choosing for $\eta$ some sausage we find a single positive eigenvalue denoting an unstable mode. The existence and the interpretation of this unstable mode will be considered in the next section.

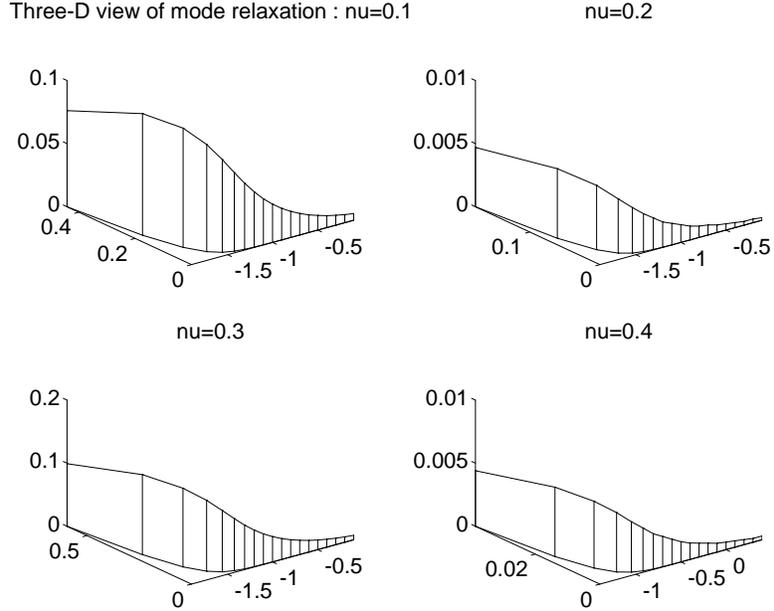

Fig.1

## 2.1 Taming the instability

It is readily verified that we have two *exact* solutions of the Jacobi variational equations *at one loop*:

$$\partial_\tau \chi = \frac{1}{2}\left(e^{-\eta}\mathbf{\Delta}_0 + e^{-\tau}\tilde{R}(\eta)\right)\chi \tag{15}$$

$$\chi_1 = \tilde{R}(\eta) \tag{16}$$

$$\chi_2 = 1 - \tfrac{1}{2}e^{-\tau}\tilde{R} \tag{17}$$

Their existence is not surprising: since we have a one–parameter set of solutions, the sausages, the derivatives with respect to $\tau$ and $\nu$ give rise to two independent solutions of the variational equations. Now, $\chi_1$ which comes from the time derivative represents the unstable solution; this kind of instability is of no concern, since it corresponds to a simple redefinition of the initial time parameter and it can be fixed by restricting to a given initial area. The other solution is tangent to the manifold of sausages. The component of $\chi$ transversal to both $\chi_1$ and $\chi_2$ represents the distance of a generic solution from the sausage manifold.

Numerical results (Tab.3 and 4) show that the transversal part is exponentially decreasing with a slope ($\approx 8.73$) much higher than the slope of $\chi_2$ itself ($\approx 1.88$). The point is that there is an overall convergence in the infrared toward the constant curvature metric. All sausages converge to a ball with vanishing radius. The rate of this convergence is however slower than the rate of decay of all other modes. The figure shows how the longitudinal component (plotted in the vertical direction) is still large when the transversal one is already negligible.

Conversely, the evolution in the ultraviolet direction $\tau \to -\infty$ is strongly unstable, which makes it quite hard to follow numerically: all truncation errors are chaotically amplified and the calculation becomes rapidly unreliable.

## 3 The geometry made manifest (embedding the sausage in $R^3$)

The evolution of the metric according to the RG equation can be easily visualized in the case of cylindrical symmetry. The problem is to recover the embedding of the surface in $R^3$ in such a way that the induced metric coincides with $\exp(\phi)\delta_{ij}$.

Let us denote by $X, Y, Z$ the Cartesian coordinates in $R^3$; let $Z$ be the axis of symmetry; we can therefore reduce the problem to the determination of the $X, Z$ section; the embedding is given by a map $y :\to [X, Z]$, $y$ being the argument of the conformal factor $\phi$. The length of a section $y = constant$ is given on one hand by $2\pi \exp(\phi/2)$ on the other by $2\pi X$, which gives one condition. The other condition comes by considering the length element at $\varphi = constant$ which yields

$$\mathrm{d}X^2 + \mathrm{d}Z^2 = \exp(\phi)dy^2$$

The two conditions combined are easily reduced to a simple quadrature which can be implemented in MATLAB. As an example, fig.2 reports the evolution of a sausage at one–loop: App.C contains some details of the algorithm.

The fast convergence towards an asymptotic sausage is depicted in Fig.3: the initial surface is a sausage with a large symmetric deformation starting at $\tau = -3$. Notice that the scale has been compressed in the Z direction in order to emphasize the dynamics of the surface; the shape at $\tau \approx 2.6$ is actually similar to that of the previous figure.

## 4 Conclusions

The RG equation of non–linear sigma–models are represented by non–linear partial differential equations, possibly non–local if the beta–function were known to all loops. While it will always be possible to find some solutions or even families of solutions in closed form, it is very unlikely that a general solution will ever be found in analytic form; the numerical approach, even within its limitations, is the only general method which can can provide useful quantitative information on the solutions. In view of the possibly non–local character of the equations and because of the fact that we have to work with a local chart on a Riemannian manifold, the spectral approach seems to be the best candidate in order to build an efficient algorithm. Our implementation, which can be easily adapted to the toroidal case by exploiting FFT, is hopefully generalizable to the totally asymmetric case. It would be nice to be able to design a *fast* algorithm analogous to the finite Fourier case, which would allow for a finer mesh.

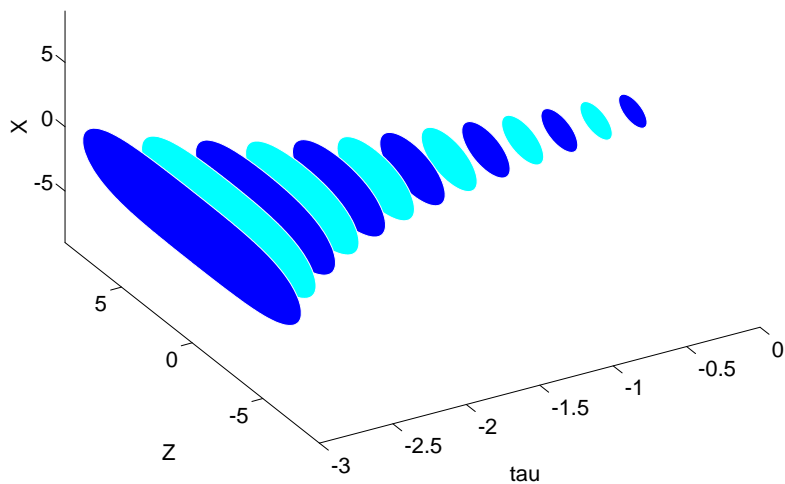

Fig.2

# Appendices

## A. Gaussian integration coefficients

The following MATLAB module [3] builds the relevant parameters for the Gaussian integration with $n$ points. The returned variables are:

1. $x$: a vector with the zeroes of the $n$-th Legendre Polynomial;

2. $p$: a matrix whose rows are given by the Legendre Polynomials $P_l(x_k)$ for $0 \leq l < n$;

3. $w$: a vector with the Gaussian integration coefficients

4. $Lapl$: a matrix representing the Laplace operator on the sphere; by construction $p(j,:) * Lapl = -j(j+1)p(j,:)$.

---

[3] We give only a subset of MATLAB modules in these appendices; the interested reader can download the whole set rgeq.uu at the WWW home page http://www.fis.unipr.it in the subsection "papers".

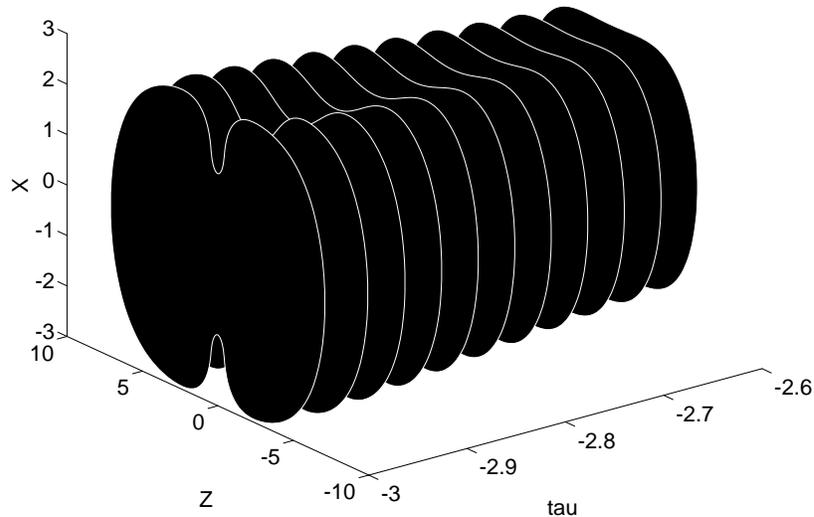

Fig.3

Since the Laplacean is obtained through its spectral representation on Legendre polynomials, in principle we get an exact representation up to polynomials of order $n$; the truncation errors however reduce the accuracy by a factor of order $n(n+1)$.

```
function [x,p,w,lapl,pder]=legendre(n)
% Legendre polynomials: zeroes and gaussian integration
% coefficients. Usage:
% [x,p,w,lapl,pder]=legendre(n)
%
disp('Building the companion matrix of P_n(x)');
l=0:n-1;
l1=1+1./(2.*l+1.);
l2=1-1./(2. *l+3.);
lp=0.5*sqrt(l1.*l2);
D=diag(lp,1)+diag(lp,-1);
disp('              ... and diagonalizing it:');
[A,B]=eig(D);
[x,ind]=sort(-diag(B)');
A=A(:,ind);
disp('Build Legendre polynomials and Gauss weights');
m=0:n;
lam=diag(1./sqrt(m+0.5));
p0=lam*A;
```

```
w=p0(1,:).^2;
p=lam*A/diag(p0(1,:));
disp('        ... and finally build the Laplace operator');
lapl=-(diag(p0(1,:))*A')*diag(m.*(m+1))*(A/diag(p0(1,:)));
j=0:n;
j1=1:n;
j1=j1.*(1.5:n+0.5);
j=j.*(j+0.5);
pder=diag(w)*p'*(diag(j)*p*diag(x)+diag(j1,-1)*p);
```

◁ After invoking legendre($n$) the expansion in Legendre polynomials of some pre-defined function *myfunc()* is a matter of two lines of code, namely:

$$\begin{aligned} \boldsymbol{f} &= \text{myfunc}(\boldsymbol{x}); \\ \boldsymbol{\mathcal{L}f} &= (\boldsymbol{f}.*\boldsymbol{w})*\boldsymbol{p}'; \\ \boldsymbol{f} &= ((N-1/2).*\boldsymbol{\mathcal{L}f})*\boldsymbol{p}; \end{aligned} \qquad (18)$$

where $N = 1 : n$. The second line implements the integral

$$(\mathcal{L}f)_l = \int_{-1}^{1} f(x)\, P_l(x)\, \mathrm{d}x$$

while the third gives the expansion

$$f(x) = \sum_{l=0}^{n-1} \frac{2l+1}{2}\, (\mathcal{L}f)_l\, P_l(x)$$

## B: main routine

To start the evolution of a surface, we have first of all to initialize the "legendre" environment by a call to *rgini.m*; we then define the initial conformal factor by selecting the shifted field $\eta$. At this point we may call the main routine *rgeq.m* based on the adaptive step Runge-Kutta routine *ode45.m* of MATLAB. Here is one implementation:

```
% rgini.m
disp('Preparing the environment for rg equation:');
disp('Please enter the number of grid points');
n=input('npts=');
global  XL WL LAPLACE PDER  PL
[XL,PL,WL,LAPLACE,PDER]=legendre(n-1);
% end

function [fout,t,devs]=rgeq(eta,ti,tf,nu,nsplit,loop,tol)
```

```
% rgeq.m
% usage: [fout,t,devs]=rgeq(eta,ti,tf,nu,nsplit,loop,tol)
% renormalization group equation for non-linear sigma model
%        eta   = initial conformal factor
%   ti   = starting time,
%    tf   = final time,
%         nsplit = number of time intervals,
%   loop =1, 2
%   tol  = acc (default 1.e-6)
% Requires global variables built by "rgini"

global XL WL PL
% normalize to fixed volume
eta=eta-log(sum(exp(eta).*WL)/2);
t0=ti;
dt=(tf-ti)/nsplit;
tc=ti+dt;
conf=[eta 0]; % conf stores also the accumulated time parameter
        % according to dt = -dtau/betaf()
N=length(conf);
x=[ti nu];
x=fmins('fitsaus',x,[0,1.e-6],[],eta,ti);
dev=norm(eta-saus(x));
fout=zeros(nsplit+1,N);
fout(1,:)=conf;
devs=[ti x dev];
% preparing to  plot  the surface
[X,Z]=surf_par(eta,ti);
fill3(ti*ones(size(X)), X , Z,'w');
hold on
drawnow
for it=1:nsplit,
if loop==1,
[tauc,f]=ode45('rgdif1',t0,tc,conf,tol); %one loop
else
[tauc,f]=ode45('rgdif2',t0,tc,conf,tol); %two loop
end
conf=f(length(tauc),:);
taueff = -log(exp(-ti)-conf(N)/(2*pi));
[X,Z]=surf_par(conf(1:N-1),tc);
fill3(tc*ones(size(X)), X , Z, 'w');
drawnow
```

```
x=fmins('fitsaus',x,[0,1.e-6],[],conf(1:N-1),tc);
dev=norm(conf(1:N-1)-saus(x));
disp([tc x dev]);
devs=[devs ; [tc  x dev]];
t0=t0+dt;
tc=tc+dt;
fout(it+1,:)=conf; % store results
end

t=2*pi*exp(-ti)-fout(:,N);
fout=fout(:,1:N-1);

%% rgdif1.m
function fprime = rgdif1(tau,eta)
fprime=zeros(size(eta));
n=length(eta);
f=eta(1:n-1);
R = curv(f,tau);
fprime(1:n-1)=1-betaone(R)/betaone(2*exp(tau));
fprime(n)= -1/betaone(2*exp(tau));
```

◁

## C: surface embedding

The following MATLAB function accepts a conformal factor $\phi = \exp(f)$ at some scale $\tau$ and returns the section of the surface embedded in $R^3$:

```
%% surface parameterization

function [X,Z]=surf_param(f,tau);

global XL PDER % define by rgini
theta=acos(XL);
dpsi= f*PDER;
dZ = -exp(0.5*(f-tau)).*sqrt(1-(XL+0.5*dpsi).^2);
X = sin(theta).*exp(0.5*(f-tau));
Z = prim(dZ,theta);
Z = [Z fliplr(Z) Z(1)];
X = [X -fliplr(X) X(1)];
%end
```

```
%%   primitive function

function p=prim(f,x)
dx=[0 diff(x)];
df=[0 diff(f)];
p=cumsum((f-0.5*df).*dx);
```

# Acknowledgments

E.O. would like to thank warmly V. A. Fateev, who introduced him to the rich world of two–dimensional models and suggested several stimulating problems. We thank warmly C. Destri for his generous collaboration.

Table 1: Evolution of an ellipsoid; $D$ is the distance to the sausage manifold.

| $\tau$ | $\nu_{eff}$ | $D$ |
|---|---|---|
| -4.00000 | 0.03761 | 0.04265 |
| -3.90000 | 0.03609 | 0.01453 |
| -3.80000 | 0.03557 | 0.00551 |
| -3.70000 | 0.03536 | 0.00217 |
| -3.60000 | 0.03528 | 0.00088 |
| -3.50000 | 0.03524 | 0.00056 |
| -3.40000 | 0.03523 | 0.00015 |
| -3.30000 | 0.03522 | 0.00006 |
| -3.20000 | 0.03522 | 0.00003 |
| -3.10000 | 0.03522 | 0.00001 |
| -3.00000 | 0.03521 | 0.00000 |
| -2.90000 | 0.03521 | 0.00000 |
| -2.80000 | 0.03521 | 0.00000 |
| -2.70000 | 0.03521 | 0.00000 |
| -2.60000 | 0.03521 | 0.00000 |
| -2.50000 | 0.03521 | 0.00000 |

Table 2: The *sausageness* of an ellipsoid with eccentricity $\epsilon$ and starting time $\tau = -3$.

| $\epsilon$ | 0.10 | 0.20 | 0.30 | 0.40 | 0.50 | 0.75 |
|---|---|---|---|---|---|---|
| $\nu_{eff}$ | 0.0037 | 0.0075 | 0.0115 | 0.0159 | 0.0210 | 0.0405 |

Table 3: Decay of longitudinal component ($\chi_{\parallel}$) and of the transversal one ($\chi_{\perp}$) at $\nu = .1$

| $\tau$ | $\chi_{\parallel}$ | $\chi_{\perp}$ |
|---|---|---|
| -1.9000 | 0.0047 | 0.1772 |
| -1.8000 | 0.0055 | 0.0734 |
| -1.7000 | 0.0051 | 0.0303 |
| -1.6000 | 0.0043 | 0.0125 |
| -1.5000 | 0.0036 | 0.0051 |
| -1.4000 | 0.0030 | 0.0021 |
| -1.3000 | 0.0025 | 0.0009 |
| -1.2000 | 0.0020 | 0.0004 |
| -1.1000 | 0.0017 | 0.0001 |
| -1.0000 | 0.0014 | 0.0001 |
| -0.9000 | 0.0011 | 0.0000 |
| -0.8000 | 0.0009 | 0.0000 |
| -0.7000 | 0.0008 | 0.0000 |
| -0.6000 | 0.0006 | 0.0000 |
| -0.5000 | 0.0005 | 0.0000 |

Table 4: Decay of longitudinal component ($\chi_{\parallel}$) and of the transversal one ($\chi_{\perp}$) at $\nu = .3$

| $\tau$ | $\chi_{\parallel}$ | $\chi_{\perp}$ |
|---|---|---|
| -1.9000 | 0.0761 | 0.4415 |
| -1.8000 | 0.0963 | 0.2057 |
| -1.7000 | 0.0944 | 0.0939 |
| -1.6000 | 0.0848 | 0.0421 |
| -1.5000 | 0.0734 | 0.0186 |
| -1.4000 | 0.0623 | 0.0081 |
| -1.3000 | 0.0524 | 0.0035 |
| -1.2000 | 0.0438 | 0.0015 |
| -1.1000 | 0.0364 | 0.0006 |
| -1.0000 | 0.0302 | 0.0003 |
| -0.9000 | 0.0250 | 0.0001 |
| -0.8000 | 0.0207 | 0.0000 |
| -0.7000 | 0.0171 | 0.0000 |
| -0.6000 | 0.0140 | 0.0000 |
| -0.5000 | 0.0116 | 0.0000 |